\documentclass[aps,prl,amssymb,showpacs,amsmath,superscriptaddress,preprintnumbers,reprint]{revtex4-1}

\usepackage{natbib}
\usepackage{soul}

\usepackage{graphicx}
\usepackage{tikz}
\usetikzlibrary{snakes}
\usepackage{color}
\definecolor{Blue}{rgb}{0.00, 0.00, 1.00}
\definecolor{Red}{rgb}{1.00, 0.00, 0.00}

\newcommand{\rme}{{\mathrm{e}}}
\newcommand{\rmd}{{\mathrm{d}}}

\newcommand{\nn}{\nonumber}
\newcommand {\Eq}[1]{Eq.\hspace{0.55ex}(\ref{#1})}

\newcommand{\fig}[2]{\includegraphics[width=#1]{./#2}}
\newcommand{\Fig}[1]{\includegraphics[width=8.7cm]{./#1}}

\renewcommand{\epsilon}{\varepsilon}

\def\be{\begin{equation}}
\def\ee{\end{equation}}

\def\bal{\begin{align}}
\def\eal{\end{align}}

\def\bea{\begin{eqnarray}}
\def\eea{\end{eqnarray}}
\renewcommand{\log}{\ln }

\usepackage{color}

\begin{document}

\title{Generalized arcsine laws for   fractional Brownian motion}

\author{Tridib Sadhu}
\affiliation{Tata Institute of Fundamental Research, Mumbai 400005, India.}
\author{Mathieu Delorme}
\author{Kay J\"org Wiese}
\affiliation{
Laboratoire de Physique Th\'eorique, D\'epartement de Physique de l'ENS, Ecole Normale Sup\'erieure, 24 rue Lhomond, 75005 Paris, France.\\ CNRS; PSL Research University; UPMC Univ.\ Paris 6, Sorbonne Universit\'es.
}

\begin{abstract}
The {\em three arcsine laws} for  Brownian motion are a cornerstone of extreme-value statistics. For a Brownian   $B_t$ starting from the origin, and evolving during time $T$, one considers the following three observables: (i) the duration $t_+$ the process is positive, (ii) the  time $t_{\rm last}$ the process last visits the origin, and (iii) the time $t_{\rm max}$ when it achieves its maximum (or minimum). All three observables have the same cumulative probability distribution expressed as an arcsine function, thus the name of {\em arcsine laws}. We show how these laws change for   fractional Brownian motion $X_t$, a non-Markovian Gaussian process indexed by the Hurst exponent $H$. It generalizes standard Brownian motion ({\em i.e.}\ $H=\tfrac{1}{2}$).  We obtain the three probabilities using a perturbative  expansion in $\epsilon = H-\tfrac{1}{2}$. While all three probabilities are different, this distinction can only be made at  second order in $\epsilon$. Our results are confirmed to high precision by extensive numerical simulations. 
\end{abstract}

\pacs{05.40.Jc, 02.50.Cw, 87.10.Mn}

\maketitle

\begin{figure}[t]
\Fig{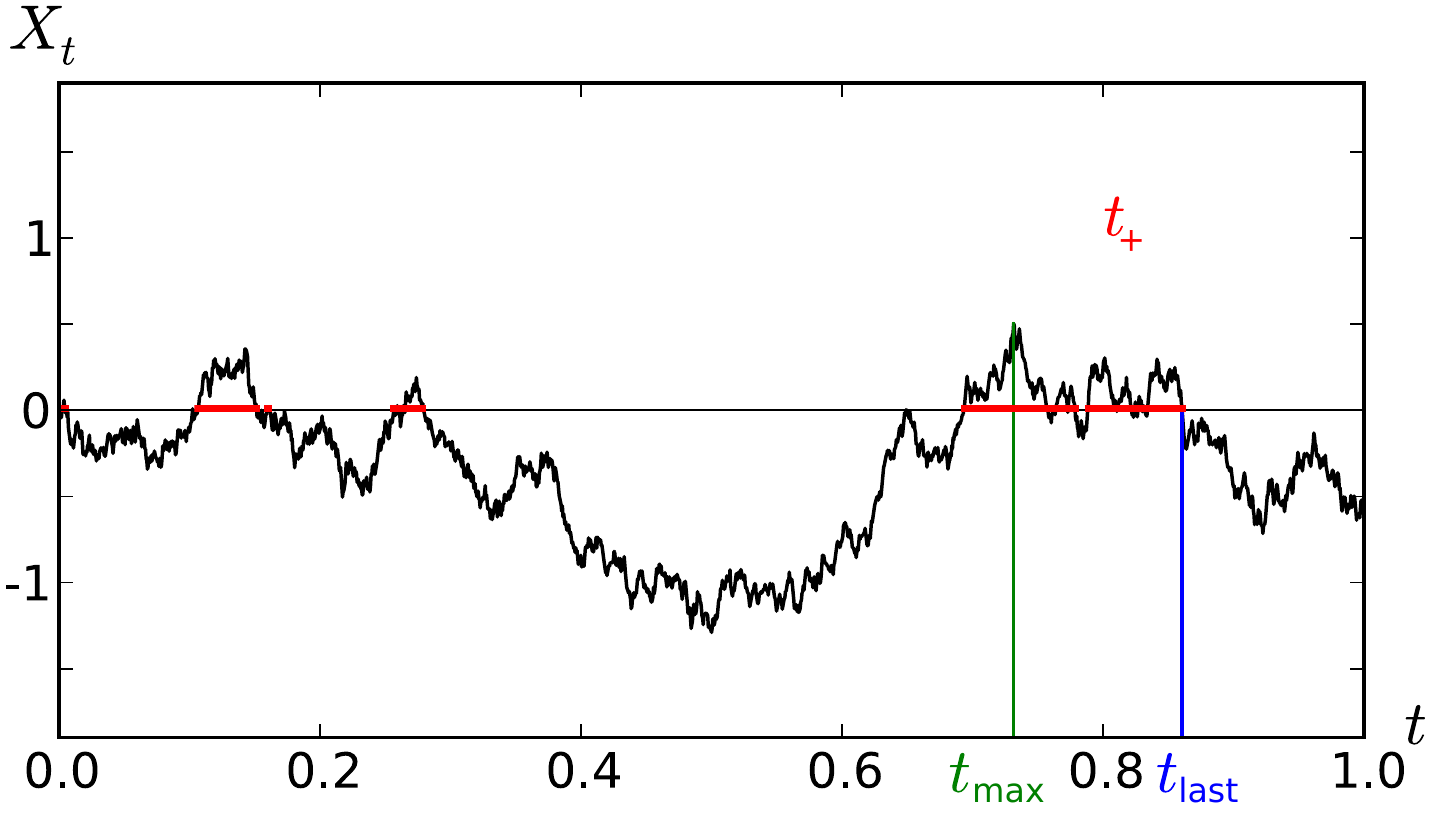}
\caption{The three observables $t_+$, $t_{\rm last}$, and $t_{\rm max}$ considered in this  letter.}
\label{f:traj}
\end{figure}

The three arcsine laws for   Brownian motion or more generally for discrete random processes \cite{Levy1940,FellerBook,Morters2010,Yen2013}
 are   celebrated properties of stochastic processes.
 For a Brownian   $B_t$ starting from the origin, and evolving during time $T$, one considers the following three observables (see Fig.~\ref{f:traj}): (i) the total duration $t_+$ when the process is positive, (ii) the last time $t_{\rm last}$ the process visits the origin, and (iii) the time $t_{\rm max}$ it achieves its maximum (or minimum). Remarkably, all three observables have the same probability distribution as a function of $\vartheta := t/T$, 
\be\label{Levey-law}
p( \vartheta)= \frac{1}{{\pi \sqrt{\vartheta(1-\vartheta)}} }\ .
\ee 
As the cumulative distribution contains an arcsine function, these laws are commonly referred to as the {\it first, second} and {\it third} arcsine law. These laws apply quite generally to   Markov processes, {\em i.e.}\ processes where the increments are uncorrelated \cite{FellerBook}. Their   counter-intuitive form with a divergence at $\vartheta=0$ and $\vartheta=1$ has sparked a lot of interest, and they are considered  among the most important properties of stochastic processes. Recent studies   led to many extensions,  in constrained Brownian motion \cite{Majumdar2008,Bouchaud2008,Randon2007}, for general stochastic processes \cite{Majumdar2010,Schehr2010,HOCHBERG1994,Pitman1992,Carmona1994,Lamperti1958}, even in higher dimensions \cite{BARLOW1989,Bingham1994,Ernst2017}. 
The laws are realized in a  plethora of real-world examples, from finance \cite{Dale1980,Baz2004} to competitive team sports \cite{Clauset2015}.

In this {\it letter}, we ask how these laws change for   fractional Brownian motion (fBm) which is a generalization of   standard Brownian motion preserving scale invariance as well as translation invariance, both in time and space. FBm was introduced in its final form by Mandelbrot and Van Ness \cite{MandelbrotVanNess1968} to describe time-series data in natural processes. It is defined as a Gaussian process $X_t$, starting at zero, $X_{0}=0$, with  mean $\left<X_{t}\right>=0$  and  covariance
\begin{equation}
\langle X_t X_s \rangle = t^{2H} + s^{2H} - |t-s|^{2H} \ . \label{eq:covariance}
\end{equation}
The parameter $H\in (0,1)$ is the Hurst exponent. 
Standard Brownian motion corresponds to $H=\tfrac{1}{2}$ where the covariance   reduces to $\langle X_t X_s \rangle = 2 \min(s,t) $. 
Unless $H=\tfrac{1}{2}$, the process is non-Markovian, \textit{i.e.}\ its increments are not independent.
For $H>\tfrac{1}{2}$ they are  positively correlated, while for $H<\tfrac{1}{2}$ they are anti-correlated. This non-Markovian nature makes a theoretical analysis of fBm difficult, and only few exact results are available in the literature \cite{Molchan1999,KrugKallabisMajumdarCornellBraySire1997,
Guerin2016}.

FBm is important as it successfully models  a variety of natural processes \cite{Decreusefond98}: a tagged particle in the single-file ($ H\,{=}\,0.25 $) \citep{KrapivskyMallickSadhu2015,SadhuDerrida2015}, the integrated current in diffusive transport ($ H\,{=}\,0.25 $) \cite{SadhuDerrida2016}, polymer translocation through a narrow pore ($ H\,{\simeq}\,0.4 $) \cite{ZoiaRossoMajumdar2009,DubbeldamRostiashvili2011,PalyulinAlaNissilaMetzler2014}, anomalous diffusion \cite{BouchaudGeorges1990}, values of the log return of a stock  ($H\,{\simeq}\,0.6\; {\rm to }\; 0.8 $) \cite{Peters1996,CutlandKoppWillinger1995,Biagini2008,Sottinen2001}, hydrology ($H\,{\simeq}\,0.72\;{\rm to}\;0.87 $) \cite{MandelbrotWallis1968}, a tagged monomer in a polymer   ($ H\,{=}\,0.25$) \cite{GuptaRossoTexier2013}, solar flare activity ($H\,{\simeq}\,0.57\;{\rm to}\;0.86$) \cite{Moreno2014}, the price of electricity in a liberated market ($ H\,{\simeq}\,0.41 $) \cite{Simonsen2003}, telecommunication networks ($H\,{\simeq}\,0.78\;{\rm to}\;0.86$) \cite{Norros2006}, telomeres inside the nucleus of human cells ($H\,{\simeq}\,0.18\;{\rm to}\;0.35$) \cite{Burnecki2012}, or diffusion inside crowded fluids ($ H\,{\simeq}\,0.4 $) \cite{Ernst2012}. Generalizing the  three arcsine laws (\ref{Levey-law}) to fBm thus  has fundamental importance, as well as a multitude of potential applications.

\begin{figure}[t]
\fig{8.5cm}{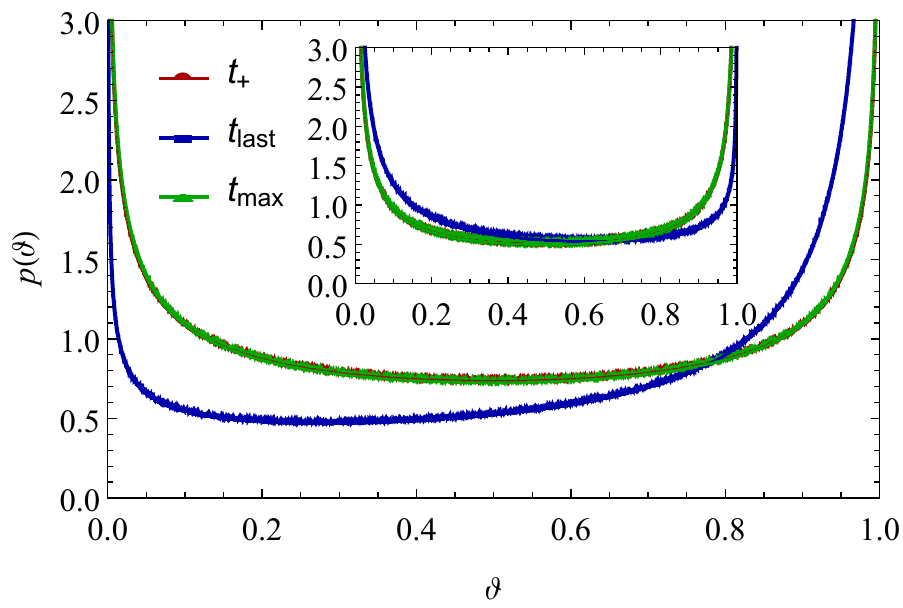}
\caption{Numerical simulation results for the probability of the three observables $t_{\rm last}$, $t_+$, and $t_{\rm max}$ for a fBm with $H=0.33$. The inset shows the probabilities for $H=0.66$. Note that the distributions of $t_+$ and $t_{\rm max}$ are almost indistinguishable.\label{f:raw}}
\end{figure}

Unlike for   Brownian motion, the probabilities of the three observables $t_+$, $t_{\rm last}$ and $t_{\rm max}$ are different. Using an expansion in $\epsilon=H-\tfrac{1}{2}$, we derive them in the   form:
\bea \label{P+}
p_+(\vartheta) &=& \frac{{\cal N}_+}{[\vartheta(1-\vartheta)]^H}\, \rme^{\epsilon {\cal F}^+_1(\vartheta) +\epsilon^2 {\cal F}^+_2(\vartheta) + {\cal O}(\epsilon^3) } \\
\label{Plast}
p_{\rm last}(\vartheta)&=&\frac{{\cal N}_{\rm last}}{\vartheta^H(1-\vartheta)^{1-H}} \,\rme^{\epsilon {\cal F}^{\rm last}_1(\vartheta) +\epsilon^2 {\cal F}^{\rm last}_2(\vartheta)  + {\cal O}(\epsilon^3) } ~~~\\
p_{\rm max}(\vartheta)&=&\frac{{\cal N}_{\rm max}}{[\vartheta(1-\vartheta)]^H}\, \rme^{\epsilon {\cal F}^{\rm max}_1(\vartheta) +\epsilon^2 {\cal F}^{\rm max}_2(\vartheta) + {\cal O}(\epsilon^3) }
\label{Pmax}
\eea
The pre-factors of the exponential are  predicted using scaling arguments for $\vartheta\to 0$ and $\vartheta\to 1$. The terms in the exponential are non-trivial, 
and finite over the full range of $\vartheta$.
We use the convention that the integral over each ${\cal F}$-function vanishes, which   adjusts the normalization constants ${\cal N}$.
To   leading order we find 
\begin{eqnarray}
\mathcal{F}_1^{+} (\vartheta)&=& \mathcal{F}_1^{\rm max} (\vartheta) \\
&=& 
\pi  \sqrt{\frac{1-\vartheta }{\vartheta }}+\frac{2 (2 \vartheta -1)\, \mbox{acos} (\sqrt{\vartheta
   } )}{\sqrt{(1-\vartheta ) \vartheta }}-\frac{\pi ^2}{2}+2 \ ,~ \nn \\
\mathcal{F}_1^{\rm last} (\vartheta)&=&  0\ .  \label{8}
\end{eqnarray}
The   expression of $\mathcal{F}_1^{\rm max} (\vartheta)$ was reported earlier \cite{DelormeWiese2015,DelormeWiese2016,DelormeWiese2016b,DelormeThesis}.
The equality of   $\mathcal{F}_1^{+}$ and $\mathcal{F}_1^{\rm max}$, and their difference from the vanishing $\mathcal{F}_1^{\rm last}$ is qualitatively seen in Fig.~\ref{f:raw}. 
We have no intuitive understanding of this   coincidence.

\begin{figure}[t]
\fig{8.5cm}{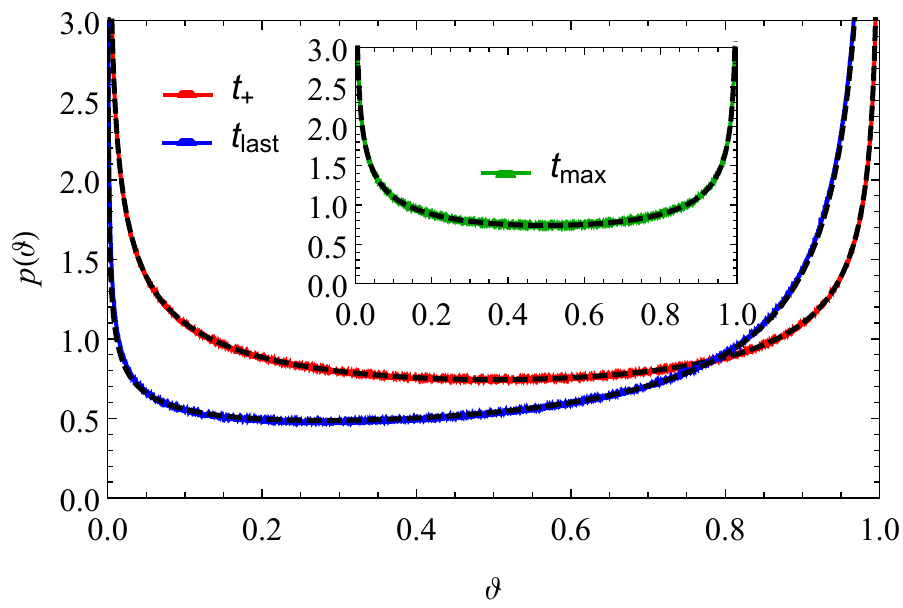}
\caption{Comparison of the formulas \eqref{P+}-\eqref{Pmax} with their corresponding numerical simulation result of a fBm with $H=0.33$. The dashed lines are the theoretical result.   $p_{\rm max}(\vartheta)$ is shown in the inset as it is almost indistinguishable from $p_{+}(\vartheta)$.}
\label{f:all}
\end{figure}

A numerical estimation of the three probabilities is obtained using a discrete-time algorithm \cite{DiekerPhD} for fBm of a given  $H$, which generates sample trajectories 
 drawn from a Gaussian probability with covariance \eqref{eq:covariance}. The probabilities in figures \ref{f:raw} and \ref{f:all} are obtained by averaging over $5\times 10^9$ sample trajectories, each  with $2^{13}$ time steps.

Figure \ref{f:raw} shows that $p_{\rm last}(\vartheta)$ behaves markedly differently from the other two distributions; especially, it is asymmetric under the exchange $\vartheta\to 1-\vartheta$.  This can be seen in the scaling part of Eq.~(\ref{Plast}), where the exponent $H$ comes from the return probability to the starting point, while the survival exponent $\theta = 1-H$ governs the divergence for $\vartheta \to 1$. 
This  asymmetry in exponents
is reversed around $H=\tfrac{1}{2}$,  as seen in the inset of figure \ref{f:raw}. 

\begin{figure*}
\newcommand{\figsize}{5.9cm}\fig{\figsize}{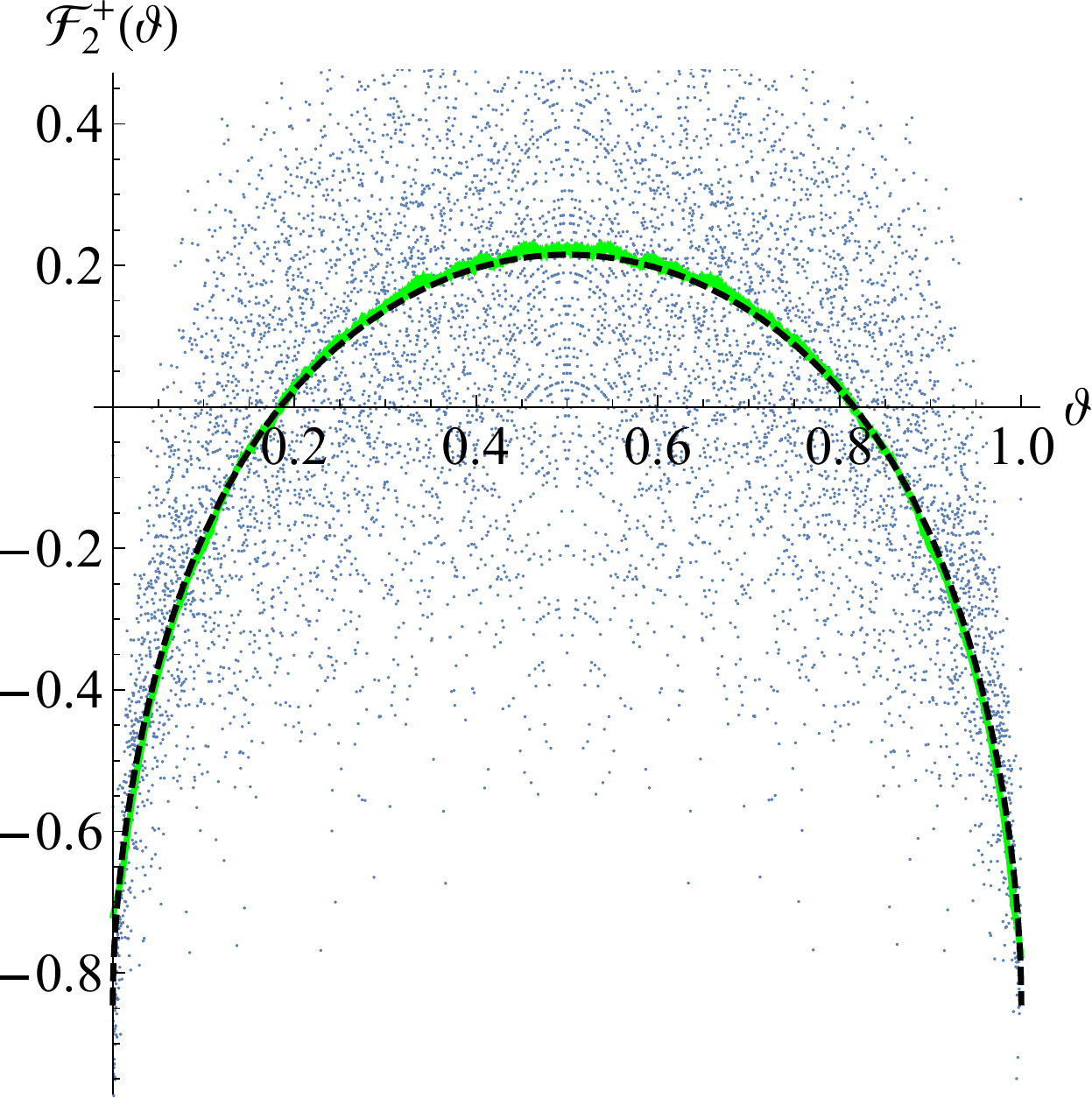} \fig{\figsize}{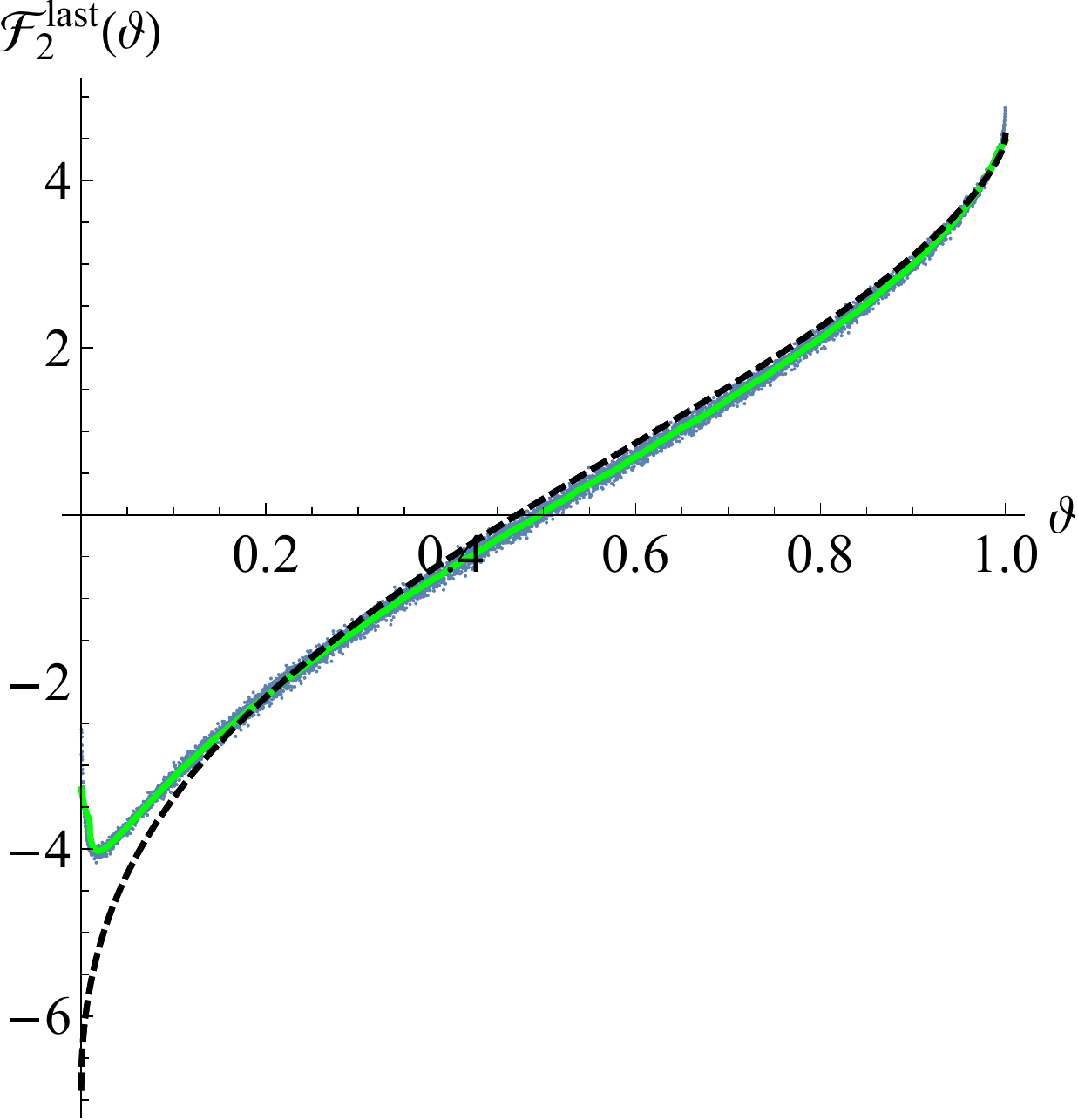}
\fig{\figsize}{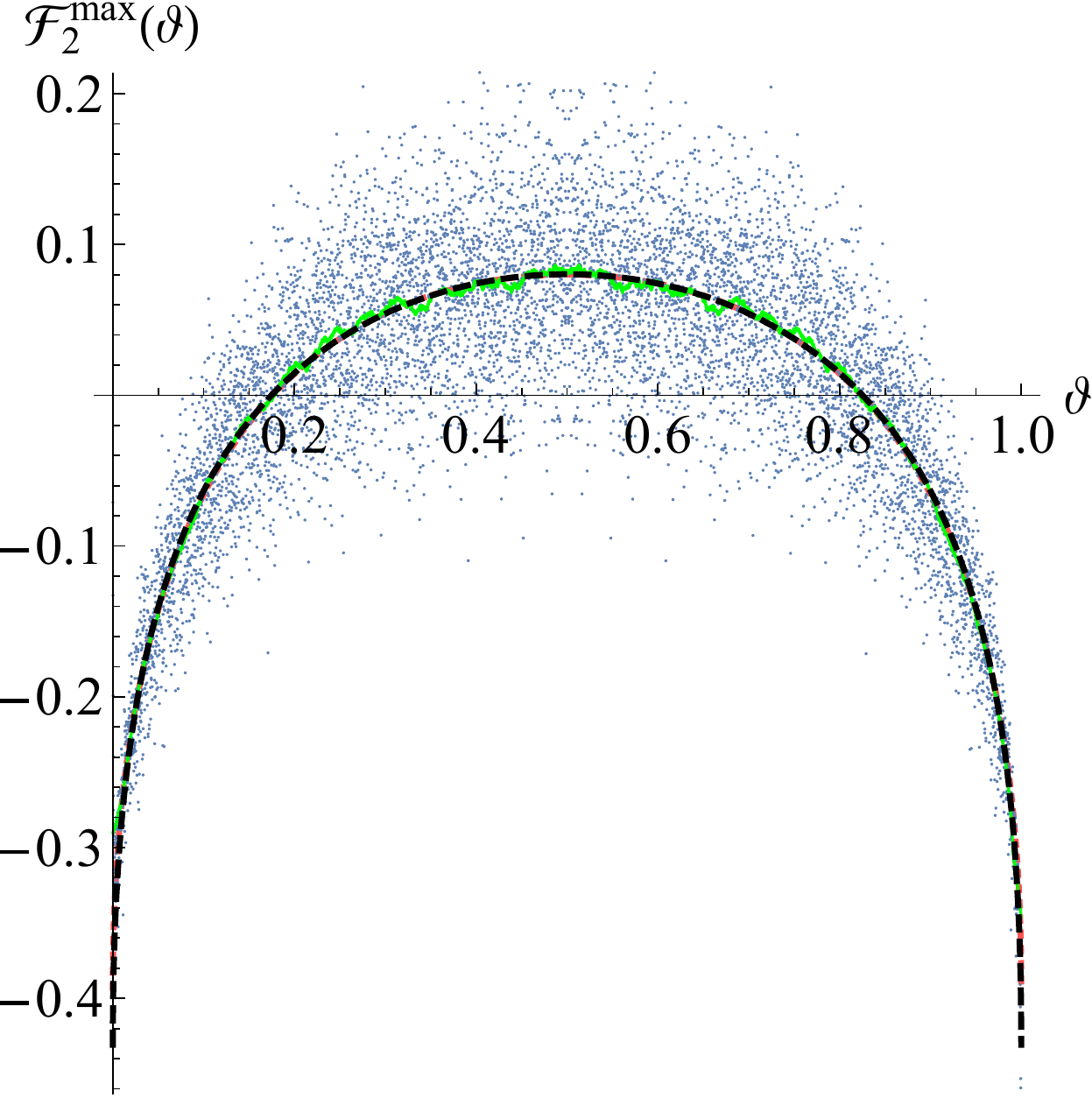} 
\put(-425,30){$(a)$}
\put(-250,30){$(b)$}
\put(-85,30){$(c)$}
\caption{A comparison  for the three  ${\cal F}_2(\vartheta)$ obtained analytically (black dashed lines) and their measurement using formula (\ref{9})   with $\epsilon=\pm\tfrac{1}{6}$. From left to right: $(a)$ positive time, $(b)$ time of the last visit to the origin, and $(c)$ time for the maximum. The scattered dots are the raw data  from trajectories of $N=2^{13}$ time steps,  averaged over $5 \times 10^9$ samples, which are coarse grained by a factor of 100 to give the green curve. Approximations of our analytical results are given in the supplementary material. \label{f:F2s}}
\end{figure*}

\begin{figure}
\newcommand{\figsize}{8.5cm} \fig{\figsize}{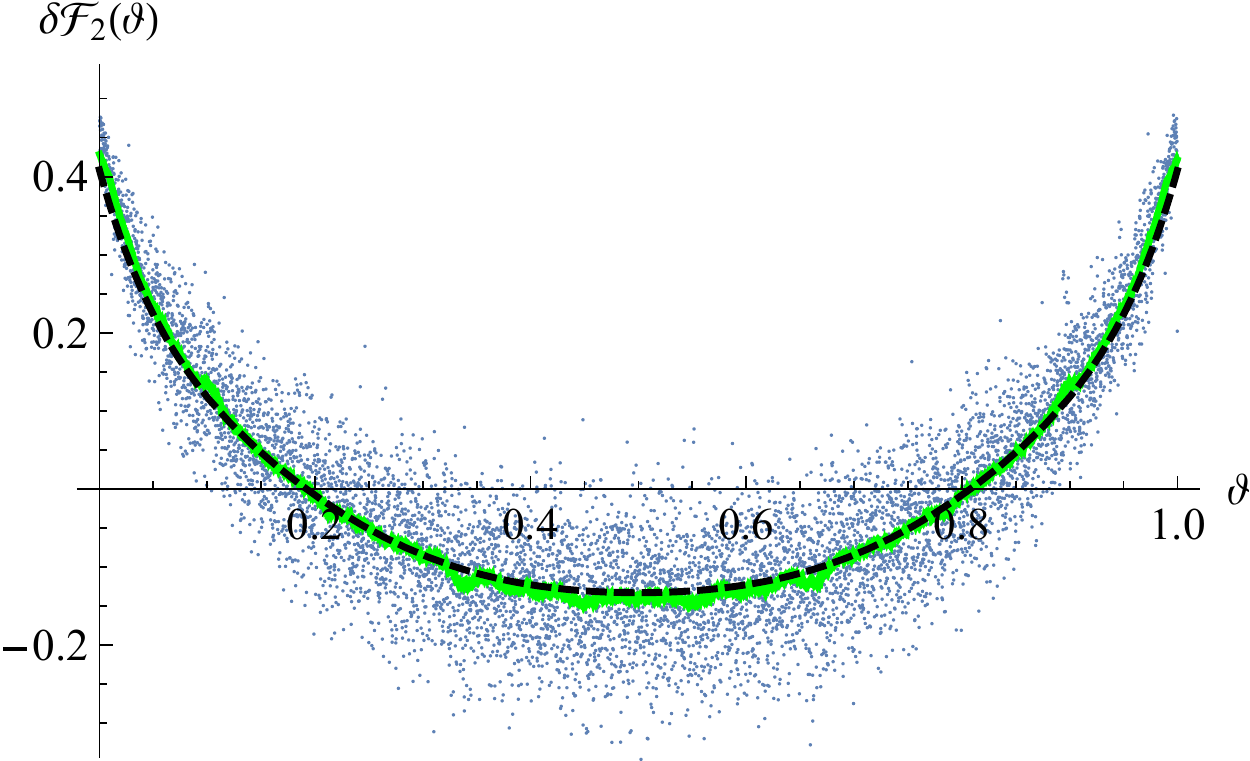} 
\caption{The difference $\delta {\cal F}_2(\vartheta) = {\cal F}_2^{\rm max}(\vartheta)-{\cal F}_2^+(\vartheta)$, using the same conventions as in Fig.~\ref{f:F2s}. This plot quantifies the difference between the first and third arcsine law.}
\label{f:F2diffs}
\end{figure}

The analytical expressions for $\mathcal{F}_2$ in Eqs.~\eqref{P+}--\eqref{Pmax} are cumbersome; we will sketch the derivation for the simplest one, ${\cal F}_2^{\rm last}(\vartheta)$, below, while the remaining ones  will be reported elsewhere \cite{SadhuWieseToBePublished}. 

Confirmation of our theoretical results comes from comparison with numerical simulations of the probabilities presented in Fig.~\ref{f:all}. For a finer comparison we plot our theoretical results of ${\cal F}_2(\vartheta)$ in Fig.~\ref{f:F2s} alongside their  extraction from numerical simulations.
To illustrate our procedure, we use Eq.~\eqref{P+} to define 
\be
{\cal F}_{2,\epsilon}^+(\vartheta):= \frac1\epsilon\left[\frac1\epsilon \ln \!\bigg( p_+(\vartheta)  \frac{[\vartheta (1-\vartheta)]^H}{\cal N^+}  \bigg) -{\cal F}_1^+(\vartheta)\right] \ .
\ee
Then, ${\cal F}_{2,\epsilon}^+(\vartheta)  = {\cal F}_{2}^+(\vartheta) + {\cal O}(\epsilon)$ which contains all terms in the exponential  in Eq.~\eqref{P+} except   ${\cal F}_{1}^+(\vartheta)$. One can improve this estimation   by using that the sub-leading term in ${\cal F}_{2,\epsilon}^+(\vartheta)$ is odd in $\epsilon$, to define
\be\label{9}
\overline{\cal F}_{2,\epsilon}^+(\vartheta):=\frac12\Big[ 
{\cal F}_{2,\epsilon}^+(\vartheta) +{\cal F}_{2,-\epsilon}^+(\vartheta)  \Big] = {\cal F}_{2}^+(\vartheta) +{\cal O}(\epsilon^2)\ .
\ee
A comparison of $\overline{\cal F}_{2,\epsilon}^+(\vartheta)$ extracted from   numerical simulations   of $p_+(\vartheta)$ with the theoretical result of ${\cal F}_{2}^+(\vartheta)$ is plotted in Fig.\ \ref{f:F2s} for $\epsilon =\pm \frac16$. The figure also  contains the comparison for   ${\cal F}_{2}^{\rm last}(\vartheta)$ and ${\cal F}_{2}^{\rm max}(\vartheta)$.
As one sees, the agreement between   theory and   numerical simulations is quite striking:  remind that these are sub-sub leading corrections, almost indiscernable in Fig.\ \ref{f:all}. We   note the much larger amplitude  of ${\cal F}_2^{\rm last}(\vartheta)$.  The latter also has the  largest deviations from the theory, especially    for $\theta \to 0$.
These deviations   indicate the presence of sub-leading terms of order $\epsilon^4$, or higher.

In Fig.\ \ref{f:raw} the probabilities $p_+(\vartheta)$ and $p_{\rm max}(\vartheta)$ are hard to distinguish from each other. Their difference can analytically be seen only at   second order in $\epsilon$. To underline  that these are  distinct distributions, we show the difference    $\delta \mathcal{F}_2(\vartheta) = {\cal F}_{2}^{\rm max}(\vartheta)-{\cal F}_{2}^{+}(\vartheta)$      in Fig.~\ref{f:F2diffs}. 

In the rest of this letter we sketch the derivation of   formulas \eqref{P+}--\eqref{Pmax}. We begin with the action which characterizes the probability of a fBm trajectory, 
\be
{\cal S}[X] =  \int_{0}^{T}\rmd t_1\int_{t_1}^{T}\rmd t_2 \;\dot X_{t_1} {\cal C}^{-1}(t_1,t_2) \dot X_{t_2}\ .
\ee
Here ${\cal C}(t_1,t_2)$ is the co-variance given in Eq.~\eqref{eq:covariance}. We use an expansion \cite{WieseMajumdarRosso2010,DelormeWiese2016} of the action around $H=\tfrac{1}{2}$ to take advantage of the Markov property of Brownian motion. One writes
\begin{align}
&{\cal C}(t_1,t_2) \nn \\
&= 2 D_{\epsilon} \left[ \delta (t_1-t_2)  +\frac{ \epsilon}{\left| t_1-t_2\right| }+\frac{2 \epsilon^2  \log  {\left| \frac{t_1-t_2}\tau \right|} 
    }{\left| t_1-t_2\right|
   } +{\cal O}(\epsilon ^3)  \right] 
\end{align}
which leads to an expansion of the action  \cite{Note1}
\bea \label{12}
  S[  X] &=& \frac1{2 D_{\epsilon}} \int_{0}^{T}\!\!\rmd t_1\int_{t_1}^{T}\!\! \rmd t_2 \;   \dot X_{t_1}\dot X_{t_2}  \Big[\delta(t_1-t_2)+ \frac{ \epsilon}{| t_1-t_2| } \nn\\
  & & \qquad  +\int_{t_1}^{t_2}\rmd s\,\frac{ \epsilon^2   
    }{| t_1-s| \,|t_2-s|
   } + {\cal O} (\epsilon ^3 )   \Big]  
\eea
where 
$
D_{\epsilon}\simeq (1+ 2\epsilon )\tau^{2\epsilon}
$ 
and all expressions are regularized by an ultraviolet cutoff $\tau$ in time.  
 
Our calculation for the probabilities is done in Laplace variables. One reason for this choice is that the space integrals appearing in   perturbation theory are easier. A further advantage   is that   temporal convolutions  become mere products in the conjugate Laplace-variables. The action \eqref{12} is written in Laplace variables \cite{DelormeWiese2016}  using
\be
\frac{  \Theta(|t_1-t_2|>\tau)}{|t_1-t_2|} \simeq \int_0^\Lambda \rmd y \, \rme^{-y|t_1-t_2|}\ ,
\ee
where $\Theta(x)$ is the Heaviside  function, and the UV cutoff $\Lambda$ 
is related to $\tau$ by $\ln( \tau) =-\ln( \Lambda )- \gamma_{\rm E}$ with $\gamma_{\rm E}$    the Euler constant.
As an explicit example, let us consider the calculation of     $P_{\rm last}(t,T)$ and its Laplace transform
\be\label{17}
\widetilde P_{\rm last}(\lambda,s)=\int_{0}^{\infty}\rmd T\int_0^{T}\rmd t\, \rme^{-\lambda t -s T}P_{\rm last}(t,T)\ .
\ee
The quantity of interest is $p_{\rm last}(\tfrac{t}{T})= \tfrac{1}{T}P_{\rm last}(t,T)$ which yields
\be\label{18}
\widetilde P_{\rm last}(\lambda,s) = \frac1s \int_0^1 \rmd \vartheta \,\frac{p_{\rm last}(\vartheta)}{1+\kappa \vartheta } \ ; \quad \kappa = \frac\lambda s\ .
\ee
Defining $\widetilde p_{\rm last}(\kappa):= s   \widetilde P_{\rm last}(\lambda,s)$, one obtains the probability $p_{\rm last}(\vartheta)$ by taking the inverse transformation
\be
\label{eq:19}  
p_{\rm last}(\vartheta) = \lim_{\phi\to \pi} \frac{-1}{\pi {\vartheta}}{\Im} \, \widetilde p_{\rm last}(\kappa = \rme^{i \phi}/\vartheta)\ , 
\ee
where $\Im$ denotes the imaginary part. This is  proven from Eq.~\eqref{18} via analytical continuation.

The calculation is simplest at order zero in $\epsilon$, {\it i.e}, for a Brownian. Using Eqs.~\eqref{12} and \eqref{17} one writes 
\be \label{157}
\widetilde P_{\rm last}^{{H=\frac{1}{2}}}(\lambda,s) =\lim_{x_{0}\to 0}\frac {2}{x_0} \int_{0}^{\infty}\!\!\!\!\!\rmd x\, \widetilde Z(x_0,x_0,s+\lambda) \widetilde Z^+(x_0,x,s)\ .
\ee
Here
$\widetilde Z(x,y,s)= (2\sqrt{s})^{-1}\exp(-\sqrt{s} |x-y|)$
is the Laplace transform of the Brownian propagator, 
while $\widetilde Z^+(x,y,s) =Z(x,y,s)-Z(x,-y,s)$ is
the propagator in presence of an absorbing wall at the origin. This yields
\be\label{17bis}
\widetilde p_{\rm last}^{H=\frac{1}{2}}(\kappa)=s\;\widetilde P_{\rm last}^{{H=\frac{1}{2}}}(s \kappa,s) =\frac{1}{\sqrt{1+\kappa}}\ .
\ee
Using the transform \eqref{eq:19} one obtains the arcsine law \eqref{Levey-law}.

\begin{figure}[t]
~~~{{\begin{tikzpicture}
\draw [->,thick] (0,0) -- (3.5,0);
\draw [->,thick] (0,0) -- (0,3);
\draw [ultra thick] (1.5,0) -- (3,0);
\draw [dashed] (1.5,0) -- (1.5,3);
\draw [dashed] (3,0) -- (3,3);
\draw [dashed] (0.5,0) -- (0.5,3);
\draw [dashed] (1,0) -- (1,3);
\node (a) at (0.6,2.6) {$\hspace{-6.4mm}(a)$};
\node (m1) at  (0,0.2) {$\hspace{-5mm}x_0$};
\node (x1) at (0.5,0.7) {$\hspace{-5mm}x_1$};
\node (x2) at (1.,1.9) {$\hspace{-5mm}x_2$};
\node (x0) at (1.5,0.2) {$\hspace{-5mm}x_0$};
\node (m2) at  (3,2.3) {$\hspace{-6mm}\parbox{0mm}{$~~~~~~~x$}$};
\draw [snake=snake,red,thick] (x1) -- (x2);
\draw [blue] (m1) -- (x1);
\draw [blue] (x1) -- (x2);
\draw [blue] (x2) -- (x0);
\draw [blue] (x0) -- (m2);
\fill (m1) circle (1.5pt);
\fill (x0) circle (1.5pt);
\fill (x1) circle (1.5pt);
\fill (x2) circle (1.5pt);
\fill (m2) circle (1.5pt);
\node (t1) at (1.5,-0.2) {$t$};
\node (t2) at (3,-0.2) {$T$};
\end{tikzpicture}}}~~~~~~~~{{\begin{tikzpicture}
\draw [->,thick] (0,0) -- (3.5,0);
\draw [->,thick] (0,0) -- (0,3);
\draw [ultra thick] (1.5,0) -- (3,0);
\draw [dashed] (1.5,0) -- (1.5,3);
\draw [dashed] (3,0) -- (3,3);
\draw [dashed] (1.9,0) -- (1.9,3);
\draw [dashed] (2.5,0) -- (2.5,3);
\node (b) at (0.6,2.6) {$\hspace{-6.4mm}(b)$};
\node (m1) at  (0,0.2) {$\hspace{-6mm}x_0$};
\node (x0) at (1.5,0.2) {$\hspace{-5mm}{}^{{}^{\textstyle x_0}}$};
\node (x3) at (1.9,2.5) {$\hspace{-4.5mm}x_1$};
\node (x4) at (2.5,1.2) {$\hspace{-6mm}x_2$};
\node (m2) at  (3,2.3) {$\hspace{-6mm}\parbox{0mm}{$~~~~~~~x$}$};
\draw [snake=snake,red,thick] (x3) -- (x4);
\draw [blue] (m1) -- (x0);
\draw [blue] (x0) -- (x3);
\draw [blue] (x3) -- (x4);
\draw [blue] (x4) -- (m2);
\fill (m1) circle (1.5pt);
\fill (x0) circle (1.5pt);
\fill (x3) circle (1.5pt);
\fill (x4) circle (1.5pt);
\fill (m2) circle (1.5pt);
\node (t1) at (1.5,-0.2) {$t$};
\node (t2) at (3,-0.2) {$T$};
\end{tikzpicture}}}

~~~~{{\begin{tikzpicture}
\draw [->,thick] (0,0) -- (3.5,0);
\draw [->,thick] (0,0) -- (0,3);
\draw [ultra thick] (1.5,0) -- (3,0);
\draw [dashed] (1.5,0) -- (1.5,3);
\draw [dashed] (3,0) -- (3,3);
\draw [dashed] (0.5,0) -- (0.5,3);
\draw [dashed] (1,0) -- (1,3);
\draw [dashed] (1.9,0) -- (1.9,3);
\draw [dashed] (2.5,0) -- (2.5,3);
\node (m1) at  (0,0.2) {$\hspace{-6mm}x_0$};
\node (c) at (0.6,2.6) {$\hspace{-6.4mm}(c)$};
\node (x1) at (0.5,0.7) {$\hspace{-5mm}x_1$};
\node (x2) at (1.,1.9) {$\hspace{-5mm}x_2$};
\node (x0) at (1.5,0.2) {$\hspace{-5mm}x_0$};
\node (x3) at (1.9,2.9) {$\hspace{-5mm}x_3$};
\node (x4) at (2.5,1.9) {$\hspace{-5mm}x_4$};
\node (m2) at  (3,2.3) {$\hspace{-6mm}\parbox{0mm}{$~~~~~~~x$}$};
\draw [snake=snake,red,thick] (x1) -- (x4);
\draw [snake=snake,red,thick] (x2) -- (x3);
\draw [blue] (m1) -- (x1);
\draw [blue] (x1) -- (x2);
\draw [blue] (x2) -- (x0);
\draw [blue] (x0) -- (x3);
\draw [blue] (x3) -- (x4);
\draw [blue] (x4) -- (m2);
\fill (m1) circle (1.5pt);
\fill (x0) circle (1.5pt);
\fill (x1) circle (1.5pt);
\fill (x2) circle (1.5pt);
\fill (x3) circle (1.5pt);
\fill (x4) circle (1.5pt);
\fill (m2) circle (1.5pt);
\node (t1) at (1.5,-0.2) {$t$};
\node (t2) at (3,-0.2) {$T$};
\end{tikzpicture}}}~~~~~~~~{{\begin{tikzpicture}
\draw [->,thick] (0,0) -- (3.5,0);
\draw [->,thick] (0,0) -- (0,3);
\draw [ultra thick] (1.5,0) -- (3,0);
\draw [dashed] (1.5,0) -- (1.5,3);
\draw [dashed] (3,0) -- (3,3);
\draw [dashed] (0.5,0) -- (0.5,3);
\draw [dashed] (1,0) -- (1,3);
\draw [dashed] (1.9,0) -- (1.9,3);
\draw [dashed] (2.5,0) -- (2.5,3);
\node (m1) at  (0,0.2) {$\hspace{-6mm}x_0$};
\node (d) at (0.6,2.6) {$\hspace{-6.5mm}(d)$};
\node (x1) at (0.5,1.4) {$\hspace{-5mm}x_1$};
\node (x2) at (1.,0.5) {$\hspace{-5mm}x_2$};
\node (x0) at (1.5,0.2) {$\hspace{-5mm}x_0$};
\node (x3) at (1.9,2.9) {$\hspace{-5mm}x_3$};
\node (x4) at (2.5,1.9) {$\hspace{-5mm}x_4$};
\node (m2) at  (3,2.3) {$\hspace{-6mm}\parbox{0mm}{$~~~~~~~x$}$};
\draw [snake=snake,red,thick] (x1) -- (x3);
\draw [snake=snake,red,thick] (x2) -- (x4);
\draw [blue] (m1) -- (x1);
\draw [blue] (x1) -- (x2);
\draw [blue] (x2) -- (x0);
\draw [blue] (x0) -- (x3);
\draw [blue] (x3) -- (x4);
\draw [blue] (x4) -- (m2);
\fill (m1) circle (1.5pt);
\fill (x0) circle (1.5pt);
\fill (x1) circle (1.5pt);
\fill (x2) circle (1.5pt);
\fill (x3) circle (1.5pt);
\fill (x4) circle (1.5pt);
\fill (m2) circle (1.5pt);
\node (t1) at (1.5,-0.2) {$t$};
\node (t2) at (3,-0.2) {$T$};
\end{tikzpicture}}}

\caption{The diagrams (a) and (b) contributing  to the order-$\epsilon$  term in $p_{\rm last}(\vartheta)$, as well as (c) and (d) contributing to the order-$\epsilon^2$  term $ {\cal F}_{2}^{\rm last}(\vartheta)$ in \eqref{Plast}. Solid lines denote the Brownian propagators, with absorbing boundary conditions indicated by a bold line after time $t$. The curly lines represent the order-$\epsilon$ interaction  in \Eq{12}.} 
\label{f:diagsD1+D2last}
\end{figure}
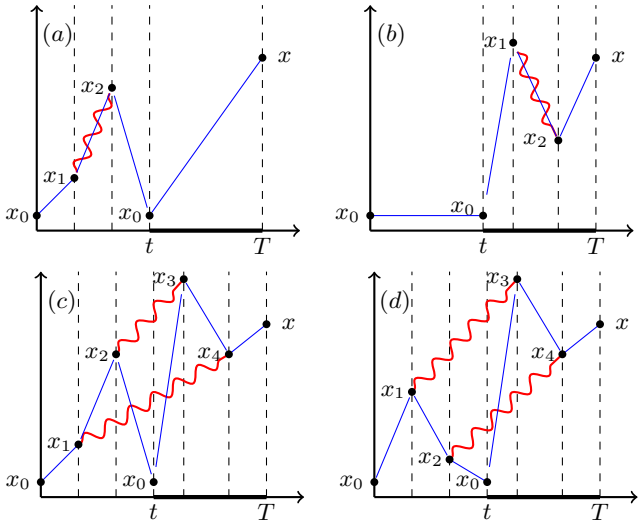%

Perturbative corrections to the probability are evaluated by following a similar procedure \cite{DelormeWiese2016}. Contributions 
at different orders in $\epsilon$ are represented by the diagrams     in Fig.~\ref{f:diagsD1+D2last}. The non-vanishing contributions at  order $\epsilon$  come from the two diagrams $(a)$ and $(b)$ which like \eqref{157} are expressed in terms of the Brownian propagator.
For example, the amplitude corresponding to   diagram $(a)$ is
\begin{align}
&  \frac {  4 \epsilon}{x_0}\int_{0}^{\Lambda} \rmd y\int_{{-\infty}}^{\infty} \rmd x_1 \int_{{-\infty}}^{\infty} \rmd x_2 \int_{0}^{\infty}\rmd x\, \widetilde Z(x_0,x_1,s_1) \\
 &~~~~~\times\partial_{x_1}\widetilde Z(x_1,x_2,s_1{+}y) \partial_{x_2}\widetilde Z(x_2,x_0,s_1) \widetilde Z^{+}(x_0,x,s)\nn
\end{align}
where $s_1= s(1+\kappa)$.
This leads to the  non-trivial power-law in \Eq{Plast} and a vanishing   ${\cal F}_1^{\rm last}$ in  \Eq{8}.

At order $\epsilon^2$, there are multiple diagrams which contribute to the probability $\widetilde p_{\rm last}(\vartheta)$. However, the only contributions to ${\cal F}_2^{\rm last}$ come   from the two diagrams $(c)$ and $(d)$ in Fig.~\ref{f:diagsD1+D2last}. After some tedious algebra, the net amplitude of the two diagrams reads
\begin{align}\label{168sym}
&\!\!\!\widetilde {\cal F}_{2}^{\rm last}(\kappa)= - \int_{0}^{\Lambda}\!\!\!\!\rmd y_1\, \rmd y_2  \bigg[ \sqrt{\kappa {+}y_1{+}y_2{+}1} -\sqrt{\kappa {+}y_1{+}1} \nn\\
   &\!\!\! ~~~~~~~~~~ -\sqrt{\kappa {+}y_2{+}1}+\sqrt{\kappa
   {+}1}\bigg] \times  \frac{ 2 \sqrt{1{+}\kappa} \sqrt{y_1{+}y_2{+}1} }{ y_1^2 y_2^2}\nn \\
&\!\!\!~~~~~~~~~~\times  \bigg(1-\sqrt{y_1{+}1}-\sqrt{y_2{+}1}+\sqrt{y_1{+}y_2{+}1}\bigg).
\end{align}
Finally,  ${\cal F}_{2}^{\rm last}(\vartheta)$ is obtained using
\be\label{23}
{\cal F}_{2}^{\rm last}(\vartheta) = \lim_{\phi\to \pi} \Re\, \widetilde {\cal F}_{2}^{\rm last}(\kappa = \rme^{i \phi}/\vartheta)\ , 
\ee
which follows from Eqs.~\eqref{eq:19} and    \eqref{Plast}   \cite{footnote2}. Integrals in Eq.\ \eqref{168sym} converge for $\Lambda \rightarrow \infty$   leading to the result shown in the middle of Fig.~\ref{f:F2s}.

Similar calculations for the other two probabilities $P_{+}(\vartheta)$ and $P_{\rm max}(\vartheta)$ are   more involved. For example, in $p_{\rm   max}(\vartheta)$ ten diagrams contribute   to the power-law $[\vartheta(1-\vartheta)]^{-H}$ in \Eq{Pmax}; in addition there are seven diagrams which contribute to ${\cal F}_2^{\rm   max}$. All these terms need to be   grouped with the appropriate repeated first-order diagrams to yield  combinations which converge for $\Lambda \to \infty$. These  calculations  will be reported   elsewhere  \cite{SadhuWieseToBePublished}.

To summarize: we   calculated the probabilities  \eqref{P+}-\eqref{Pmax}   generalizing  the three arcsine laws to fBm up to order  $\epsilon^2$, improved by incorporating the exact  scaling results for   $\vartheta \to 0$ and $1$.  Our numerical simulations confirm these highly non-trivial predictions accurately. 


Most  realizations of fBm found in practical applications fall within the range $H\simeq \tfrac{1}{2} \pm 0.3$ where our formulas yield  high-precision predictions.
Our approach  further  offers a systematic framework to obtain other analytical results for   non-Markovian processes,  of which very few are available so far. 

\acknowledgements
We thank  J.\ Klamser, P.~Krapivsky,  S.N.~Majumdar  and A.~Rosso for stimulating  discussions, and PSL for support by grant ANR-10-IDEX-0001-02-PSL. This research was supported in part by 
ICTS 
(Code: ICTS/Prog-NESP/2015/10).


\begin{thebibliography}{10}

\bibitem{Levy1940}
P.~L{\'e}vy,
\newblock {\em Sur certains processus stochastiques homog\`enes},
\newblock Compositio Mathematica {\bf 7} (1940)   283--339.

\bibitem{FellerBook}
W.~Feller,
\newblock {\em Introduction to Probability Theory and Its Applications},
\newblock John Wiley \& Sons, 1950.

\bibitem{Morters2010}
P.~M{\"{o}}rters and Y.~Peres,
\newblock {\em Brownian Motion},
\newblock Cambridge University Press, 2010.

\bibitem{Yen2013}
Ju-Yi Yen and M.~Yor,
\newblock {\em Paul L{\'e}vy's arcsine Laws},
\newblock Springer International Publishing, Cham, 2013.

\bibitem{Majumdar2008}
S.~N. Majumdar, J.~Randon-Furling, M.~J. Kearney  and M.~Yor,
\newblock {\em On the time to reach maximum for a variety of constrained
  Brownian motions},
\newblock J. Phys. A {\bf 41} (2008)   365005.

\bibitem{Bouchaud2008}
S.~N. Majumdar and J.~P. Bouchaud,
\newblock {\em Optimal time to sell a stock in the Black-Scholes model:
  comment on {{`}Thou shalt buy and hold{'}, by A. Shiryaev, Z. Xu and X. Y.
  Zhou}},
\newblock Quantitative Finance {\bf 8} (2008)   753--760.

\bibitem{Randon2007}
J.~Randon-Furling and S.~N. Majumdar,
\newblock {\em Distribution of the time at which the deviation of a {B}rownian
  motion is maximum before its first-passage time},
\newblock J. Stat. Mech. (2007)   P10008.

\bibitem{Majumdar2010}
S.~N. {Majumdar},
\newblock {\em {Universal first-passage properties of discrete-time random
  walks and {L{\'e}vy} flights on a line: Statistics of the global maximum and
  records}},
\newblock Physica A {\bf 389} (2010)   4299--4316.

\bibitem{Schehr2010}
G.~Schehr and P.~le Doussal,
\newblock {\em Extreme value statistics from the real space renormalization
  group: Brownian motion, Bessel processes and continuous time random walks},
\newblock J. Stat. Mech (2010)   P01009.

\bibitem{HOCHBERG1994}
K.~J. Hochberg and E.~Orsingher,
\newblock {\em The arcsine law and its analogs for processes governed by signed
  and complex measures},
\newblock Stoch. Proc. App. {\bf 52} (1994)   273--292.

\bibitem{Pitman1992}
J.~Pitman and M.~Yor,
\newblock {\em Arcsine laws and interval partitions derived from a stable
  subordinator},
\newblock Proc. London Math. Soc (1992)   65--326.

\bibitem{Carmona1994}
P.~Carmona, F.~Petit  and M.~Yor,
\newblock {\em Some extensions of the arcsine law as partial consequences of
  the scaling property of {B}rownian motion},
\newblock Prob. Theo. Rel. Fields {\bf 100} (1994)   1--29.



\bibitem{Lamperti1958}
J.~Lamperti,
\newblock {\em An occupation time theorem for a class of stochastic processes},
\newblock Trans. Am. Math. Soc. {\bf 88} (1958)   380--387.



\bibitem{BARLOW1989}
M.~Barlow, J.~Pitman  and M.~Yor,
\newblock {\em Une extension multidimensionnelle de la loi de l'arc-sinus},
\newblock in {\em Siminaire de Probabilites XXIII}, pages 294--314, Springer,
  Berlin, 1989.

\bibitem{Bingham1994}
N.~H. Bingham and R.~A. Doney,
\newblock {\em On higher-dimensional analogues of the arcsine law},
\newblock J. App. Prob. {\bf 25} (1988)   120--131.

\bibitem{Ernst2017}
P.~A. Ernst and L.~Shepp,
\newblock {\em On occupation times of the first and third quadrants for planar
  {B}rownian motion},
\newblock J. Appl. Prob. {\bf 54} (2017)   337--342.

\bibitem{Dale1980}
D.~Charles and W.~Rosemarie,
\newblock {\em The arcsine law and the treasury bill futures market},
\newblock Financial Analysts Journal {\bf 36} (1980)   71--74.

\bibitem{Baz2004}
J.~Baz and G.~Chacko,
\newblock {\em Financial Derivatives: Pricing, Applications, and Mathematics},
\newblock Cambridge University Press, 2004.

\bibitem{Clauset2015}
A.~Clauset, M.~Kogan  and S.~Redner,
\newblock {\em Safe leads and lead changes in competitive team sports},
\newblock Phys. Rev. E {\bf 91} (2015)   062815--062826.

\bibitem{MandelbrotVanNess1968}
B.~B. Mandelbrot and J.~W. {Van~Ness},
\newblock {\em Fractional {B}rownian motions, fractional noises and
  applications},
\newblock SIAM Review {\bf 10} (1968)   422--437.

\bibitem{Molchan1999}
G.~M. Molchan,
\newblock {\em Maximum of a fractional {B}rownian motion: Probabilities of
  small values},
\newblock Communications in Mathematical Physics {\bf 205} (1999)   97--111.

\bibitem{KrugKallabisMajumdarCornellBraySire1997}
J.~Krug, H.~Kallabis, S.~N. Majumdar, S.~J. Cornell, A.~J. Bray  and C.~Sire,
\newblock {\em Persistence exponents for fluctuating interfaces},
\newblock Phys. Rev. E {\bf 56} (1997)   2702--2712.

\bibitem{Guerin2016}
T~Gu{\'{e}}rin, N~Levernier, O~B{\'{e}}nichou  and R~Voituriez,
\newblock {\em {Mean first-passage times of non-Markovian random walkers in
  confinement}},
\newblock Nature {\bf 534} (2016)   356--359.

\bibitem{Decreusefond98}
L.~Decreusefond and A.~S. {\"{U}}st\"unel,
\newblock {\em Fractional Brownian motion: Theory and applications},
\newblock in {\em ESAIM: PROC}, pages 75--86, 1998.

\bibitem{KrapivskyMallickSadhu2015}
P.~L. Krapivsky, K.~Mallick  and T.~Sadhu,
\newblock {\em Dynamical properties of single-file diffusion},
\newblock J. Stat. Mech. (2015)   P09007,
\newblock arXiv:{\bf 1505.01287}.

\bibitem{SadhuDerrida2015}
T.~Sadhu and B.~Derrida,
\newblock {\em {Large deviation function of a tracer position in single-file
  diffusion}},
\newblock J. Stat. Mech. (2015)   P09008.

\bibitem{SadhuDerrida2016}
T.~Sadhu and B.~Derrida,
\newblock {\em {Correlations of the density and of the current in
  non-equilibrium diffusive systems}},
\newblock J. Stat. Mech. (2016)   113202.

\bibitem{ZoiaRossoMajumdar2009}
A.~Zoia, A.~Rosso  and S.N. Majumdar,
\newblock {\em Asymptotic behavior of self-affine processes in semi-infinite
  domains},
\newblock Phys. Rev. Lett. {\bf 102} (2009)   120602.

\bibitem{DubbeldamRostiashvili2011}
J.~L.~A. Dubbeldam, V.~G. Rostiashvili, A.~Milchev  and T.~A. Vilgis,
\newblock {\em Fractional {B}rownian motion approach to polymer translocation:
  The governing equation of motion},
\newblock Phys. Rev. E {\bf 83} (2011)   011802.

\bibitem{PalyulinAlaNissilaMetzler2014}
V.~Palyulin, T.~Ala-Nissila  and R.~Metzler,
\newblock {\em Polymer translocation: The first two decades and the recent
  diversification},
\newblock Soft Matter {\bf 10} (2014)   9016--9037.

\bibitem{BouchaudGeorges1990}
J.-P. Bouchaud and A.~Georges,
\newblock {\em Anomalous diffusion in disordered media: Statistical mechanisms,
  models and physical applications},
\newblock Phys. Rep. {\bf 195} (1990)   127--293.

\bibitem{Peters1996}
{E. E.} Peters,
\newblock {\em Chaos and order in the capital markets},
\newblock Wiley finance editions,
\newblock Wiley, New York, 2 edition, 1996.

\bibitem{CutlandKoppWillinger1995}
N.J. Cutland, P.E. Kopp  and W.~Willinger,
\newblock {\em Stock price returns and the {Joseph} effect: A fractional
  version of the {Black-Scholes} model},
\newblock in E.~Bolthausen, M.~Dozzi  and F.~Russo, editors, {\em Seminar on
  Stochastic Analysis, Random Fields and Applications}, {\em {\em Volume}~36}
  of {\em Progress in Probability}, pages 327--351, Birkh{\"a}user Basel, 1995.

\bibitem{Biagini2008}
F.~Biagini, Y~Hu, B.~Oksendal  and T.~Zhang,
\newblock {\em Stochastic Calculus for Fractional Brownian Motion and
  Applications},
\newblock Springer Verlag, London, 2008.

\bibitem{Sottinen2001}
T.~Sottinen,
\newblock {\em Fractional Brownian motion, random walks and binary market
  models},
\newblock Finance and Stochastics {\bf 5} (2001)   343--355.

\bibitem{MandelbrotWallis1968}
B.B. Mandelbrot and J.R. Wallis,
\newblock {\em Noah, {J}oseph, and operational hydrology},
\newblock Water Resources Research {\bf 4} (1968)   909--918.

\bibitem{GuptaRossoTexier2013}
S.~Gupta, A.~Rosso  and C.~Texier,
\newblock {\em Dynamics of a tagged monomer: Effects of elastic pinning and
  harmonic absorption},
\newblock Phys. Rev. Lett. {\bf 111} (2013)   210601.

\bibitem{Moreno2014}
E.~Monte-Moreno and M.~Hern{\'{a}}ndez-Pajares,
\newblock {\em Occurrence of solar flares viewed with GPS: Statistics and
  fractal nature},
\newblock J. Geophys. Res. {\bf 119} (2014)   9216--9227.

\bibitem{Simonsen2003}
I.~Simonsen,
\newblock {\em Measuring anti-correlations in the nordic electricity spot
  market by wavelets},
\newblock Physica A {\bf 322} (2003)   597--606.

\bibitem{Norros2006}
I.~Norros,
\newblock {\em On the use of fractional Brownian motion in the theory of
  connectionless networks},
\newblock IEEE J. Sel. A. Commun. {\bf 13} (2006)   953--962.

\bibitem{Burnecki2012}
K.~Burnecki, E.~Kepten, J.~Janczura, I.~Bronshtein, Y.~Garini  and A.~Weron,
\newblock {\em Universal algorithm for identification of fractional Brownian
  motion. A case of telomere subdiffusion},
\newblock Biophysical Journal {\bf 103} (2012)   1839--1847.

\bibitem{Ernst2012}
D.~Ernst, M.~Hellmann, K.~J\"{u}rgen  and M.~Weiss,
\newblock {\em {Fractional Brownian motion in crowded fluids}},
\newblock Soft Matter {\bf 8} (2012)   4886--4889.

\bibitem{DelormeWiese2015}
M.~Delorme and {K.J.} Wiese,
\newblock {\em The maximum of a fractional {Brownian} motion: Analytic results
  from perturbation theory},
\newblock Phys. Rev. Lett. {\bf 115} (2015)   210601,
\newblock arXiv:{\bf 1507.06238}.


\bibitem{DelormeWiese2016}
M.~Delorme and K.J. Wiese,
\newblock {\em Perturbative expansion for the maximum of fractional {Brownian}
  motion},
\newblock Phys. Rev. E {\bf 94} (2016)   012134,
\newblock arXiv:{\bf 1603.00651}.

\bibitem{DelormeWiese2016b}
M.~Delorme and K.J. Wiese,
\newblock {\em Extreme-value statistics of fractional {Brownian} motion
  bridges},
\newblock Phys. Rev. E {\bf 94} (2016)   052105,
\newblock arXiv:{\bf 1605.04132}.

\bibitem{DelormeThesis}
M.~Delorme,
\newblock {\em Stochastic processes and disordered systems, around Brownian
  motion},
\newblock PhD thesis, PSL Research University, 2016.

\bibitem{DiekerPhD}
A.~B. Dieker,
\newblock {\em Simulation of fractional {B}rownian motion},
\newblock PhD thesis, University of Twente, 2004.

\bibitem{SadhuWieseToBePublished}
T.~Sadhu and K.J. Wiese,
\newblock to be published.

\bibitem{WieseMajumdarRosso2010}
K.J. Wiese, S.N. Majumdar  and A.~Rosso,
\newblock {\em Perturbation theory for fractional {Brownian} motion in presence
  of absorbing boundaries},
\newblock Phys. Rev. E {\bf 83} (2011)   061141,
\newblock arXiv:{\bf 1011.4807}.

\bibitem{Note1}
{The action is written in a form that when evaluating it in perturbation theory, so-called ``contact-terms'' proportional to   $\delta$ functions in time from the contraction of two $\dot X$ have to be discarded.}

\bibitem{footnote2}
It contains the real part $\Re$, as it is divided by the order-$0$ result \eqref{17bis}. 


\end{thebibliography}

\widetext
\begin{center}
\textbf{\large Supplementary Material: Result for ${\cal F}_2$}
\end{center}
\setcounter{equation}{0}
\setcounter{figure}{0}
\setcounter{table}{0}
\makeatletter
\renewcommand{\theequation}{S\arabic{equation}}
\renewcommand{\thefigure}{S\arabic{figure}}
\renewcommand{\bibnumfmt}[1]{[S#1]}
\renewcommand{\citenumfont}[1]{S#1}
The analytical expressions for  ${\cal F}_2$ are quite complicated, and contain as \Eq{168sym}   numerical integrals to be performed. In order that the reader can use our results, we give simple approximations to the results obtained after numerical integration of these integrals,   which fit the analytical formulas to a good extent.
\be
{\cal F}_2^{+}(\vartheta)\simeq -0.842235+1.76479 \left[ \vartheta(1-\vartheta)\right]^{\frac{1}{2}}+3.70810 \left[ \vartheta(1-\vartheta)\right] -9.71973 \left[ \vartheta(1-\vartheta)\right]^{\frac{3}{2}}+7.40511 \left[ \vartheta(1-\vartheta)\right]^{2}
\ee

\be
{\cal F}_2^{\rm last}(\vartheta)\simeq -17.92401+13.30207 \sqrt{\vartheta}-2.16604 \sqrt{1-\vartheta} + 8.30059 \vartheta +11.59529 \vartheta^{\frac{3}{2}}+13.23121 (1-\vartheta)^{\frac{3}{2}}-10.74274 \vartheta^2
\ee
\be
{\cal F}_2^{\rm max}(\vartheta)\simeq -0.431001+1.69259 \left[ \vartheta(1-\vartheta)\right]^{\frac{1}{2}}-1.93367 \left[ \vartheta(1-\vartheta)\right] +1.3572 \left[ \vartheta(1-\vartheta)\right]^{\frac{3}{2}}-0.33995  \left[ \vartheta(1-\vartheta)\right]^{2}
\ee
Note that ${\cal F}_2^{+}(\vartheta)$ and ${\cal F}_2^{\rm max}(\vartheta)$ are symmetric under the exchange of $\vartheta \rightarrow 1-\vartheta$ while ${\cal F}_2^{\rm last}(\vartheta)$ is not.

\end{document}